\documentclass[pra,twocolumn,showpacs]{revtex4}
\usepackage{graphicx}
\usepackage{bm}

\begin{document}

\title{Model for Entangled States with Spin-Spin Interaction}
\author{Yakir Aharonov$^{(1)(2)}$, Jeeva Anandan$^{(1)*}$, G. Jordan Maclay$^{(3)}$, and Jun Suzuki$^{(1)}$}
\affiliation{$^*$Deceased}
\affiliation{$^{(1)}$Department of Physics and Astronomy, University of South Carolina,
Columbia, SC 29208}
\affiliation{$^{(2)}$School of Physics and Astronomy, Tel Aviv University, Tel Aviv
69978, Israel}
\affiliation{$^{(3)}$Quantum Fields LLC, 20918 Wildflower Lane, Richland Center, WI 53581}
\keywords{entangled states, dipole-dipole interaction, protective measurement}
\pacs{03.65.Ud, 03.67.Lx, 03.67.Mn}

\begin{abstract}
A system consisting of two neutral spin 1/2 particles is analyzed for two
magnetic field perturbations: 1) an inhomogeneous magnetic field over all
space, and 2) external fields over a half space containing only one of the
particles. The field is chosen to point from one particle to the other,
which results in essentially a one-dimensional problem. A number of
interesting features are revealed for the first case: the singlet, which has
zero potential energy in the unperturbed case, remains unstable in the
perturbing field. The spin zero component of the triplet evolves into a
bound state with a double well potential, with the possibility of tunneling.
Superposition states can be constructed which oscillate between entangled
and unentangled states. For the second case, we show that changes in the
magnetic field around one particle affect measurements of the spin of the
entangled particle not in the magnetic field nonlocally. By using protective
measurements, we show it is possible in principle to establish a nonlocal
interaction using the two particles, provided the dipole-dipole potential
energy does not vanish and is comparable to the potential energy of the
particle in the external field.
\end{abstract}

\date{May 03, 2004}
\maketitle




\section{Introduction}

It is of fundamental interest to investigate the properties of elementary
systems in quantum physics.  We have considered a system consisting of two
neutral spin 1/2 particles that interact through the spin-spin potential. 
This distinctly quantum system displays many of the characteristic features
of quantum systems, including entanglement, tunneling, bound states,
decaying states, and spontaneous symmetry breaking.  A similar system,
consisting of harmonically trapped alkali ions that may be entangled through
the dipole-dipole potential, has been proposed for use in quantum computing%
\cite{nielsen,you,unanyan}.  Hence there is practical as well as
fundamental interest in the model we consider.  We wanted to investigate a
system of two entangled particles that interact through a potential that
vanishes at infinity.  One purpose of the model is to allow us to
investigate the behavior of one portion of an entangled system when we
adiabatically perturb the other portion, and thereby to investigate the
coupling between the separate portions of the entangled system as a function
of the distance between them.  As mentioned, the properties exhibited may
find application in quantum computing.  In our model, we find a coupling
between portions of an entangled state that in principle allows one to send
signals by the modulation of the magnetic field provided the potential
energy from the dipole potential does not vanish.  The maximum separation
possible is probably of the order of micrometers or less and depends on the
maximum modulation frequency of the signal.  In principle, protective
measurements\cite{aavprotec,aavwave} can be done in one region of the system
to determine the elements of the reduced density matrix, something which
would not be possible using conventional measurements unless an ensemble of
identical systems was available\cite{anandanden}.  The use of protective
measurements would permit the entanglement to remain.  The adiabatic
perturbation in the protective measurement can be as large as desired, so
long as the state evolves continuously and the instantaneous energy
eigenvalue does not cross that of other levels.

Other interesting features of this spin-spin coupling model are apparent
when we apply an adiabatic perturbation that is an inhomogeneous magnetic
field over all space.  We find the initial singlet evolves into an unbound
state and the triplet develops a double hump potential, suggestive of
spontaneous symmetry breaking. For the latter case, it is possible for one
particle to tunnel across the barrier to the other side.  Also, we are able
to study a system in which a superposition evolves continuously in time,
with the wavefunction changing from entangled to unentangled and back to
entangled.  When we eliminate a spatial cutoff and allow the dipole-dipole
potential to become infinite, spontaneous symmetry breaking occurs in the
degenerate ground state. 

The paper is organized as follows.  A model for entangled states via
spin-spin interaction is constructed in Sec. II.  We study this model for
an inhomogeneous magnetic field over all space in Sec. III, and for a
constant and inhomogeneous magnetic field over a half space containing only
one of the particles in Sec. IV. We discuss the possibility of the
protective measurement in our model in Sec. V.  We summarize our results in
Sec. VI.  Two appendices include another possible model and detail
calculations outlined in the text.


\section{The Model}

We will assume that we have a pair (designated 1 and 2) of identical,
uncharged, spin 1/2 particles with coordinates $\bm{x}_{1}$ and $\bm{x}_{2}$
(and corresponding momenta $\bm{p}_{1}$ and $\bm{p}_{2})$.  We will apply a
magnetic field and determine the evolution of the system in the impulsive
approximation, in which we assume the kinetic energy of the system does not
change as we apply the magnetic field.  We will consider two cases. Case 1:
an inhomogeneous magnetic field is present throughout all space, and Case 2:
the magnetic field is present only in the region to the right of the origin.

For both cases, the Hamiltonian without external fields for our system is 
\begin{equation}
H_{0}=\frac{\bm{p}_{1}^{2}}{2m}+\frac{\bm{p}_{2}^{2}}{2m}+U(|\bm{x}_{1}-%
\bm{x}_{2}|)
\end{equation}%
and the potential energy $U$ for two interacting magnetic dipoles $\mu %
\bm{\sigma}_{1}$ and $\mu \bm{\sigma}_{2}$ located at $\bm{x}_{1}$ and $%
\bm{x}_{2}$ respectively is 
\begin{equation}
U=\mu ^{2}\frac{\bm{\sigma}_{1}\cdot \bm{\sigma}_{2}-3(\bm{\sigma}_{1}\cdot %
\bm{n})(\bm{\sigma}_{2}\cdot \bm{n})}{|\bm{x_{1}}-\bm{x}_{2}|^{3}}-\frac{%
8\pi }{3}\mu ^{2}\delta (\bm{x_{1}}-\bm{x}_{2})  \label{U}
\end{equation}%
where $\bm{n}$ is a unit vector in the direction $(\bm{x}_{1}-\bm{x}_{2})$.
 The first term is the usual dipole-dipole interaction while the second
term is the hyperfine interaction term.

In our paper, we will first analyze the model with the approximation that
the changes in the kinetic energy, potential energy, and the relative
position of the particles are all negligible when we turn on a perturbing
magnetic field (impulsive approximation)\cite{messiah}.  In this
approximation, the position of the particle does not change significantly
during the interaction.  This approximation is in the same spirit as the
Born-Oppenheimer approximation in which the electronic motion about the
nuclei of a diatomic molecule is much more rapid than the vibrational motion
of the nuclei.  Thus it is possible to obtain the eigenfunction for the
nuclear motion using the energy eigenvalue for the electronic motion as the
potential.  On the other hand we assume that during the application of the
magnetic field the spins evolve adiabatically.  This approximation is based
on the observation that the spin precession for the states is much faster
than the translational motion.  With this impulsive approximation, there is
no change in the potential energy or the relative position of the particles
and in this sense the states act as if bound.  To determine if the states
are in fact bound, one would have to treat the separation as a dynamical
variable, and include the potential and kinetic energy terms in solving
Schrodinger's equation.  If the potential energy eigenvalues obtained with
the impulsive approximation are positive, then it is very unlikely that the
state is in fact bound.  Only states with negative potential energy
eigenvalues could be bound.

Under these assumptions we can choose the coordinates system so that the
particles are separated along the $z$-direction, and $z_{1}$ and $z_{2}$ have
opposite signs.  Therefore we can rotate the coordinate system such that $(%
\bm{x}_{1}-\bm{x}_{2})\rightarrow (z_{1}-z_{2})\bm{n}$ where $\bm{n}$ is a
unit vector in the $z$-direction.  With this coordinate transformations the
term $\bm{\sigma}_{i}\cdot \bm{n}$($i=1,2$) becomes $\sigma _{z_{i}}$. 
Using $\bm{\sigma}_{1}\cdot \bm{\sigma}_{2}=2\bm{S}^{2}-3$ and neglecting
the hyperfine interaction term which is relevant only at short distances
(see the discussion in Appendix A), we can approximate $U$ as 
\begin{equation}
U(|z|)=\mu ^{2}\frac{(2\bm{S}^{2}-3)-3\sigma _{z_{1}}\sigma _{z_{2}}}{|z|^{3}%
}, z=z_{1}-z_{2}.
\end{equation}%
The classical dipole force will always be in the $z$-direction, and
therefore the separation will always be along the $z$-axis, and we have
essentially a one dimensional problem.  For two classical dipoles oriented
along the $z$-axis, the interaction would result in an attractive
(repulsive) force if the dipoles were parallel (antiparallel).

We will study the effects of applying an adiabatically perturbing magnetic
field also in the $z$-direction.  With this special choice of field, the
problem remains a one dimensional problem since the magnetic force on each
particle is also in the $z$-direction.

The spin components of the eigenstates of $H_{0}$ will be simultaneous
eigenstates of the total spin $S$ and the total spin in the $z$-direction, $%
S_{z}=S_{z_{1}}+S_{z_{2}}$.  These states comprise the usual singlet state
with $S=0$, $S_{z}=0$ corresponding to $\left\vert S\right\rangle $, and the
triplet spin eigenstates with $S=1;S_{z}=-1,0,+1$ corresponding to $%
\left\vert T_{S_{z}}\right\rangle =\left\vert T_{-1}\right\rangle
,\left\vert T\right\rangle ,\left\vert T_{1}\right\rangle )$.  Thus the
dipole-dipole potential results in entangled states\cite{you}.  Because of
the indistinguishability of the particles, it is not possible to describe
which particle is on the left or right, instead quantum mechanics indicates
there is a superposition of both.  The spatial part of the wavefunctions is
chosen so the total wavefunction is antisymmetric with respect to the
interchange of particles 1 and 2. Introducing the symmetrized
(antisymmetrized) wave function by 
\begin{equation}
\psi _{\pm }(z_{1},z_{2})=\frac{1}{\surd 2}(\psi _{R}(z_{1})\psi
_{L}(z_{2})\pm \psi _{L}(z_{1})\psi _{R}(z_{2}))  \label{SA}
\end{equation}%
where $\psi _{R}(z_{i})$ ($i=1,2$) represents the wavefunction for particle $%
i$ on the right side of the origin ($z_{i}>0$), and $\psi _{L}(z_{i})$ on
the left side ($z_{i}<0$), the singlet state is 
\begin{equation}
\left\vert S\right\rangle =\frac{1}{\surd 2}(\left\vert +\right\rangle
_{1}\left\vert -\right\rangle _{2}-\left\vert -\right\rangle _{1}\left\vert
+\right\rangle _{2})\psi _{+}(z_{1},z_{2}).
\end{equation}%
The spin 1 triplet states for $S_{z}=0,+1,-1$ respectively are 
\begin{equation}
\left\vert T\right\rangle =\frac{1}{\surd 2}(\left\vert +\right\rangle
_{1}\left\vert -\right\rangle _{2}+\left\vert -\right\rangle _{1}\left\vert
+\right\rangle _{2})\psi _{-}(z_{1},z_{2})
\end{equation}%
\begin{equation}
\left\vert T_{1}\right\rangle =\left\vert +\right\rangle _{1}\left\vert
+\right\rangle _{2}\psi _{-}(z_{1},z_{2})
\end{equation}%
\begin{equation}
\left\vert T_{-1}\right\rangle =\left\vert -\right\rangle _{1}\left\vert
-\right\rangle _{2}\psi _{-}(z_{1},z_{2}).
\end{equation}%
Using the following relations 
\begin{eqnarray}
\bm{\sigma}_{1}\cdot \bm{\sigma}_{2}\left\vert S\right\rangle 
&=&-3\left\vert S\right\rangle  \\
\bm{\sigma}_{1}\cdot \bm{\sigma}_{2}\left\vert T_{i}\right\rangle 
&=&\left\vert T_{i}\right\rangle \text{ \ \ for all}\ i \\
\sigma _{z_{1}}\sigma _{z_{2}}\left\vert S\right\rangle  &=&-\left\vert
S\right\rangle  \\
\sigma _{z_{1}}\sigma _{z_{2}}\left\vert T\right\rangle  &=&-\left\vert
T\right\rangle  \\
\sigma _{z_{1}}\sigma _{z_{2}}\left\vert T_{\pm 1}\right\rangle 
&=&\left\vert T_{\pm 1}\right\rangle ,
\end{eqnarray}%
the spin-spin interaction $U$ can be represented with the basis states $%
\{\left\vert T_{-1}\right\rangle ,\left\vert T_{1}\right\rangle ,\left\vert
T\right\rangle ,\left\vert S\right\rangle \}$ as follows:

\begin{equation}
U=f(r)\left[ 
\begin{array}{cccc}
-2 & 0 & 0 & 0 \\ 
0 & -2 & 0 & 0 \\ 
0 & 0 & 4 & 0 \\ 
0 & 0 & 0 & 0%
\end{array}
\right]
\end{equation}
where we define 
\begin{equation}
f(r)=\frac{\mu ^{2}}{r^{3}},\quad r=\left\vert z_{1}-z_{2}\right\vert .
\end{equation}

The unperturbed potential energies corresponding to the eigenstates $%
\left\vert T_{-1}\right\rangle ,\left\vert T_{1}\right\rangle ,\left\vert
T\right\rangle ,\left\vert S\right\rangle $ are $-2f,$ $-2f,$ $4f,$ $0$
respectively. The states $\left\vert T_{-1}\right\rangle ,\left\vert
T_{1}\right\rangle $ could be considered as analogous to classical systems
in which the two magnetic moments are parallel to each other, leading to
attractive forces between the particles, whereas the $\left\vert
T\right\rangle ,\left\vert S\right\rangle $ states could be considered
analogous to classical systems in which the magnetic moments are
antiparallel resulting in repulsive forces.  It is interesting that the
singlet state has zero energy for the spin-spin interaction and therefore is
not expected to be a bound state.  There is no classical analog to this
unique quantum mechanical result of zero potential energy for the singlet
which follows from the properties of quantized angular momentum.  The
triplet states characterized by a negative energy are the only states that
might be bound if the kinetic energy were included.  In any event, the
separation will not change significantly as we apply the magnetic field
perturbation.

At time $t=0$ we assume we turn on the interaction Hamiltonian $H_{I}$: 
\begin{equation}
H_{I}=-\mu (\sigma _{z_{1}}B(z_{1},t)+\sigma _{z_{2}}B(z_{2},t)).
\end{equation}%
We assume that we turn on the $B$ field slowly (adiabatically) in the $z$%
-direction, so that the spin system can adjust to the new field and
therefore remain in an eigenstate of the instantaneous Hamiltonian
(adiabatic theorem) \cite{messiah}.  The total reduced Hamiltonian $H^{R}$
(neglecting kinetic energy terms) is now 
\begin{equation}
H^{R}=\mu ^{2}\frac{\bm{\sigma}_{1}\cdot \bm{\sigma}_{2}-3\sigma
_{z_{1}}\sigma _{z_{2}}}{r^{3}}-\mu (\sigma _{z_{1}}B(z_{1},t)+\sigma
_{z_{2}}B(z_{2},t))
\end{equation}%
where $B(z_{1},t)$ is the value of the magnetic field, which always points
along the $z$-direction, at $(z_{1},t).$  As noted previously, this choice
of field is done for simplicity and is not the most general field that can
be applied.  We also note that in general this field is not consistent
with the requirement that the divergence of the magnetic field vanish \cite%
{comment1}.  To meet this requirement for an inhomogeneous field in the $z$%
-direction, we need to also have a large constant magnetic field in the $z$%
-direction.  In order to determine the evolution of the initial singlet or
triplet state when the magnetic field is applied, we find it convenient to
decompose the interaction term as 
\begin{equation}
H_{I}=-\mu (\Sigma _{-}+\Sigma _{+}),  \label{pmsigma}
\end{equation}%
where we define 
\begin{eqnarray}
\Sigma _{-} &=&\frac{1}{2}(\sigma _{z_{1}}-\sigma
_{z_{2}})(B(z_{1},t)-B(z_{2},t)) \\
\Sigma _{+} &=&\frac{1}{2}(\sigma _{z_{1}}+\sigma
_{z_{2}})(B(z_{1},t)+B(z_{2},t)).
\end{eqnarray}%
There are two cases of magnetic fields that we will consider.  In the first
case, an inhomogeneous $B$ field is proportional to $z$ everywhere. In the
second case the external magnetic field is present on the right side of the
origin only ($z_{1},z_{2}>0$).


\section{Case 1: Inhomogeneous magnetic field in the $Z$-direction present
in all space}

The time independent magnetic field is defined by 
\begin{equation}
B(z_{i})=B_{0}+bz_{i},\ i=1,2  \label{B1}
\end{equation}%
where $B_{0}$ and $b$ are constants satisfying the condition mentioned
earlier.  For this choice of the magnetic field we first apply the large
constant field impulsively so that there will be no transitions among the
states.  Then we turn on the inhomogeneous part adiabatically \cite%
{commentBb}.  In this case we can show that the total Hamiltonian $%
H_{T}=H_{0}+H_{I}$ is separable into two commuting terms that depend,
respectively, on the center of mass coordinate $Z$ and the relative position
coordinate $z$.  We define the center of mass coordinate $Z=\frac{1}{2}%
(z_{1}+z_{2}),$ with conjugate center of mass momentum $P=p_{z_{1}}+p_{z_{2}}
$ and the relative position coordinate $z=z_{1}-z_{2}$, with conjugate
momentum $p=\frac{1}{2}(p_{z_{1}}-p_{z_{2}})$.  Since $[p,Z]=[P,z]=0$ and $%
[P,Z]=[p,z]=-i$, we obtain the Hamiltonian (the reduced mass is $m/2$ for
two identical particles) 
\begin{equation}
H_{0}=\frac{P^{2}}{2m}+\frac{p^{2}}{m}+U(|z|).
\end{equation}%
Using the decomposition (\ref{pmsigma}) the total Hamiltonian is rewritten 
\begin{equation}
H_{T}=H_{1}(P,Z)+H_{2}(p,z)
\end{equation}%
where the term depending on the center of mass coordinate $Z$ is 
\begin{equation}
H_{1}=\frac{P^{2}}{2m}-\mu (\sigma _{z_{1}}+\sigma _{z_{2}})(B_{0}+bZ)
\end{equation}%
and the term depending on the relative coordinate $z$ is 
\begin{equation}
H_{2}=\frac{p^{2}}{m}+U(|z|)-\mu \frac{b}{2}(\sigma _{z_{1}}-\sigma
_{z_{2}})z.
\end{equation}%
In the subspace spanned by $\left\vert S\right\rangle $ and $\left\vert
T\right\rangle $ we can show that $[\bm{\sigma}_{1}\cdot \bm{\sigma}_{2}$, $%
\sigma _{z_{1}}+\sigma _{z_{2}}]=0.$  The commutators with only terms in $%
\sigma _{z_{1}}$ and $\sigma _{z_{2}}$ vanish, so $[H_{1},H_{2}]=0$, and the
total energy is the sum of the energy eigenvalues for $H_{1}$ and $H_{2}$.

First we find that the total reduced Hamiltonian $H^{R}$ is expressed in the
basis $\{ \left\vert T_{-1}\right\rangle ,\left\vert T_{1}\right\rangle
,\left\vert T\right\rangle ,\left\vert S\right\rangle \}$ 
\begin{eqnarray}  \label{HR1}
H^{R}&=&U+H_{I} \\
&=&f(r)\left[ 
\begin{array}{cccc}
-2 & 0 & 0 & 0 \\ 
0 & -2 & 0 & 0 \\ 
0 & 0 & 4 & 0 \\ 
0 & 0 & 0 & 0%
\end{array}
\right] \\ \nonumber
&+& \left[ 
\begin{array}{cccc}
2 \mu (B_0+bZ) & 0 & 0 & 0 \\ 
0 & -2 \mu (B_0+bZ) & 0 & 0 \\ 
0 & 0 & 0 & -gz \\ 
0 & 0 & -gz & 0%
\end{array}
\right]
\end{eqnarray}
where $g=b\mu /2$ is introduced.

Next this reduced Hamiltonian can be represented in the $\left\vert
S\right\rangle $, $\left\vert T\right\rangle $ subspace as 
\begin{equation}
H_{2}^{R}=4f(r)\left[ 
\begin{array}{cc}
1 & 0 \\ 
0 & 0%
\end{array}%
\right] -gz\left[ 
\begin{array}{cc}
0 & 1 \\ 
1 & 0%
\end{array}%
\right] .
\end{equation}%
We express it in terms of the Pauli matrices 
\begin{equation}
H_{2}^{R}=2f(r)\bm{I}+\bm{\beta}\cdot \bm{\sigma }
\end{equation}%
where $\bm{I}$ is the identity operator and $\bm{\beta }=(-gz,0,2f(r))$. 
Then the eigenvalues are given in terms of the angle $\omega $ which is
defined by 
\begin{equation}
\tan \omega =\frac{\beta _{x}}{\beta _{z}}=\frac{-gz}{2f}=\frac{-bzr^{3}}{%
4\mu }
\end{equation}%
and we choose the branch 
\begin{equation}
\sin \omega =\frac{-gz}{\sqrt{g^{2}r^{2}+4f^{2}}}=\frac{-bzr^{3}}{\sqrt{%
16\mu ^{2}+b^{2}r^{8}}}.
\end{equation}%
Physically $\tan \omega $ represents the ratio of the energy of the dipole
in the external inhomogeneous magnetic field to the energy due to the
dipole-dipole coupling.  Solving for the eigenvectors and eigenvalues of $%
H_{2}^{R}$, we obtain 
\begin{eqnarray}
\left\vert -a\right\rangle  &=&-\sin \frac{\omega }{2}\left\vert
T\right\rangle +\cos \frac{\omega }{2}\left\vert S\right\rangle  \\
E_{-}^{R} &=&2f-|\bm{\beta }|=2f(1-\sec \omega )
\end{eqnarray}%
and 
\begin{eqnarray}
\left\vert +a\right\rangle  &=&\cos \frac{\omega }{2}\left\vert
T\right\rangle +\sin \frac{\omega }{2}\left\vert S\right\rangle  \\
E_{+}^{R} &=&2f+|\bm{\beta }|=2f(1+\sec \omega ).
\end{eqnarray}

If the magnetic field, which is proportional to $b$, vanishes, then $\sin
\omega \rightarrow 0$ and $\left\vert -a\right\rangle \rightarrow \left\vert
S\right\rangle $, $\left\vert +a\right\rangle \rightarrow \left\vert
T\right\rangle $, as expected. For $\left\vert -a\right\rangle $, the
effective potential $E_{-}^{R}(z)$ is negative and always concave down so no
stable singlet states are expected (Figures 1 and 2).  For $\left\vert
+a\right\rangle $ the effective potential $E_{+}^{R}(z)$ is a positive
double hump potential which suggests stable states as shown in Figures 3 and
4.  The peak near $z_{1}\approx z_{2}\approx 0$ is due to the rapid increase
in the dipole-dipole potential, and the slopes of the flat portions of $%
E_{+}^{R}$ on either side of the peak are proportional to $b$, the
derivative of the inhomogeneous magnetic field.

\begin{figure}[tbph]
\begin{center}
\includegraphics[width=3in]{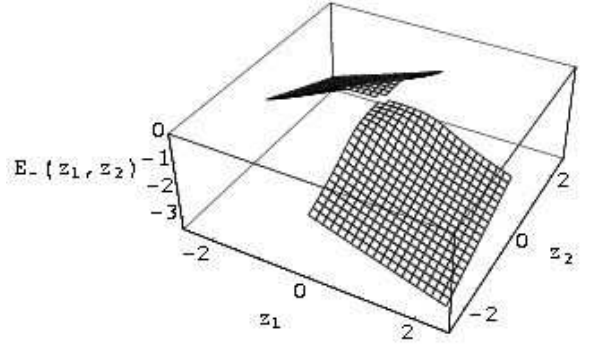}
\end{center}
\caption{\textbf{$E_{-}^{R}(z_{1},z_{2})$ for an inhomogeneous field over
all space plotted for systems in which the particles are on opposite sides
of the origin. The system could be represented by a point in the near
quadrant, in which case $z_1>0$ and $z_2<0$. If particle 1 and particle 2 are
interchanged, then the system would be represented by a point in the far
quadrant obtained by a reflection across the line $z_1=z_2$. For $E_{-}^{R}$ the
energy is monotonically decreasing as either particle moves away from the
origin and there is no stable state. All the energies are scaled by $E_{0}=%
\mu B_{0}$ and the coordinates are scaled by $r_{0}$ which is
defined by $E_{0}=2f(r_{0})\Leftrightarrow r_{0}=(2\mu /B_{0})^{1/3}$%
. }}
\label{default}
\end{figure}

\begin{figure}[htbp]
\begin{center}
\includegraphics[width=3in]{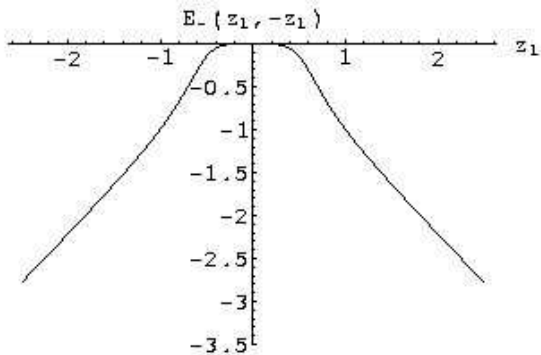}
\end{center}
\caption{\textbf{$E_- ^R(z_1,-z_1)$ for an inhomogeneous field over all
space. This is the same as the contour with $z_1+z_2=0$ in the FIG. 1.}}
\label{default}
\end{figure}

\begin{figure}[tbph]
\begin{center}
\includegraphics[width=3in]{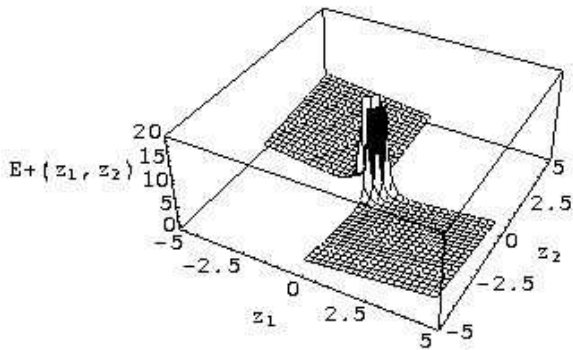}
\end{center}
\caption{\textbf{The positive eigenvalue $E_{+}^{R}(z_{1},z_{2})$ for an
inhomogeneous field over all space. We see stable states can occur in this
case. }}
\end{figure}

\begin{figure}[htbp]
\begin{center}
\includegraphics[width=3in]{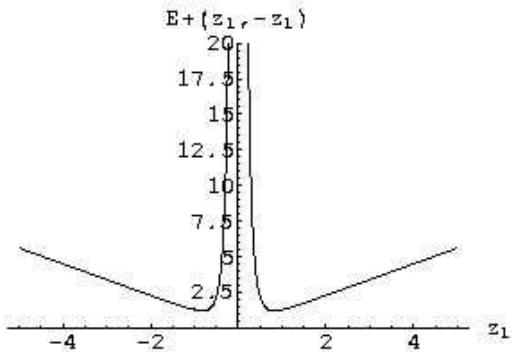}
\end{center}
\caption{\textbf{$E_+ ^R(z_1,-z_1)$ for an inhomogeneous field over all
space. The rapid increase is due to the divergence of the dipole-dipole
potential. }}
\label{}
\end{figure}

It is interesting to determine the nature of the spin wavefunctions for $%
\left\vert \pm a\right\rangle $ for large and small values of separation $r$
when a magnetic field is present.  For large $r(z>0)\Leftrightarrow
r\rightarrow \infty \ (\sin \omega \rightarrow -1)$: 
\begin{eqnarray}
\left\vert -a\right\rangle  &\rightarrow &\frac{1}{\sqrt{2}}(\left\vert
T\right\rangle +\left\vert S\right\rangle )=\left\vert +\right\rangle
_{1}\left\vert -\right\rangle _{2}, \\
E_{-}^{R} &\rightarrow &-gr, \\
\left\vert +a\right\rangle  &\rightarrow &\frac{1}{\sqrt{2}}(\left\vert
T\right\rangle -\left\vert S\right\rangle )=\left\vert -\right\rangle
_{1}\left\vert +\right\rangle _{2}, \\
E_{+}^{R} &\rightarrow &+gr
\end{eqnarray}%
and for small $r\Leftrightarrow r\rightarrow 0\ (\sin \omega \rightarrow 0):$
\begin{eqnarray}
\left\vert -a\right\rangle  &\rightarrow &\left\vert S\right\rangle
, \ E_{-}^{R}\rightarrow -\frac{b^{2}}{16}r^{5} \\
\left\vert +a\right\rangle  &\rightarrow &\left\vert T\right\rangle
, \ E_{+}^{R}\rightarrow \frac{4\mu ^{2}}{r^{3}}.
\end{eqnarray}%
For large separations, the energy eigenvalue is dominated by the effect of
the inhomogeneous field, whereas for small separations, the eigenvalue is
dominated by the dipole-dipole interaction.  As the relative position $z$ of
two particles goes from $-\infty $ to $0$ to $+\infty ,$ the state
corresponding to $\left\vert +a\right\rangle $ goes from an unentangled
state, namely $\left\vert -\right\rangle _{1}\left\vert +\right\rangle _{2}$%
, to an entangled triplet $\left\vert T\right\rangle $ state, and then back
to the unentangled $\left\vert -\right\rangle _{1}\left\vert +\right\rangle
_{2}$. By superposition of states with different momenta, 
we could form a state that would oscillate in time between 
entangled and unentangled states\cite{super}.

We need to avoid energy levels crossing each other during the application of
the adiabatic perturbation. Otherwise transitions between the levels may
occur. From the reduced Hamiltonian (\ref{HR1}) one can easily find the
energy levels for the other members of the triplets $\left\vert T_{\pm
1}\right\rangle $ to be $-2f(r)\pm 2\mu (B_{0}+bZ)$. Therefore, under the
assumption that the two particles are on the opposite sides of the origin ($%
Z\approx 0$), we require the following equalities to avoid crossing of
energy levels: 
\begin{equation}
-2f(r)+2\mu B_{0}>E_{+}^{R}>E_{-}^{R}>-2f(r)-2\mu B_{0},  \label{equality1}
\end{equation}%
which are obeyed provided the energy contribution from the constant magnetic
field $B_{0}$ is significantly greater than the contributions from either
the inhomogeneous field or the dipole potential ($\mu B_{0}>3f(r)+g|z|/2$) 
\cite{condition}.

Independently we also need a condition in order for the adiabatic evolution
to proceed (see e.g. \cite{messiah}).  Namely we need to turn on the
magnetic field (inhomogeneous part) slowly, and the condition for the time
period T for this switching is 
\begin{equation}
T\gg \frac{\hbar (gz)^{2}}{(4f)^{2}\sqrt{(gz)^{2}+(2f)^{2}}}
\end{equation}%
or, 
\begin{equation}
T\gg \frac{\hbar b^{2}r^{11}}{32\mu ^{3}\sqrt{b^{2}r^{8}+16\mu ^{2}}}.
\label{adia1}
\end{equation}%
Therefore the greater the separation, the more slowly the inhomogeneous
field needs to be applied.


\subsection{Use of Born-Oppenheimer Approximation to the Tunneling effect}

So far we have neglected the kinetic term under the assumptions that the
characteristic frequencies of the spin precession are much greater than
those of the translational motion of the two particles. These assumptions
are strictly true, for instance, for the NMR case where particles are part
of the molecules and hence they are always bound.

In contrast, we want to consider a situation where the particles are
essentially free.  For the eigenstate $\left\vert +a\right\rangle $, a graph
of the corresponding eigenvalue $E_{+}^{R}(z)$ is a positive double hump
function which describes two bound particles, one on either side of the
origin. The potential permits tunneling across the barrier. In this
tunnelling process the two particles would be exchanged. First we note
that the Schrodinger equation for the center of mass motion as well as the
relative motion is invariant under the transformation $z_{1}\rightarrow z_{2}
$ and $z_{2}\rightarrow z_{1}$ which corresponds to the tunneling
transition. In $(z_{1},z_{2})$ space, this transformation corresponds to
a reflection across the line $z_{1}=z_{2}$. The corresponding states are
degenerate in energy. The Hamiltonian is also invariant under the parity
transformation $z_{1}\rightarrow -z_{1}$, $z_{2}\rightarrow -z_{2}$.

In order to analyze this process, we can utilize a Born-Oppenheimer
approximation and we can separate two degrees of freedom by the use of an
average potential in the Schrodinger equation for the relative motion of the
particles. Thus the problem is reduced to a one body problem with the
potential for the relative coordinates. The appropriate potential is in fact
the energy eigenvalue $E_{+}^{R}(z)$ which was obtained in the previous
section: 
\begin{equation}
V(z)=E_{+}^{R}(z)=\frac{2\mu ^{2}}{|z|^{3}}+\sqrt{g^{2}z^{2}+\frac{4\mu ^{4}%
}{z^{6}}}.  \label{VB}
\end{equation}

Therefore this approximation yields: 
\begin{eqnarray}  \label{schBO}
H_{BO} \left\vert E \right\rangle& =& E\left\vert E\right\rangle \\
H_{BO} &=& \frac{p^{2}}{m}+E_{+}^{R}(z).  \label{HBO}
\end{eqnarray}

As is usual for the singular potential $\propto r^{-n}$ ($n>2$) the barrier
becomes infinitely high at the origin. To estimate the tunneling probability
we introduce a cutoff for small $r$ defined by $r_{c}$. Figure 5 shows a
typical shape of the potential barrier after this regularization.

\begin{figure}[htbp]
\begin{center}
\includegraphics[width=3in]{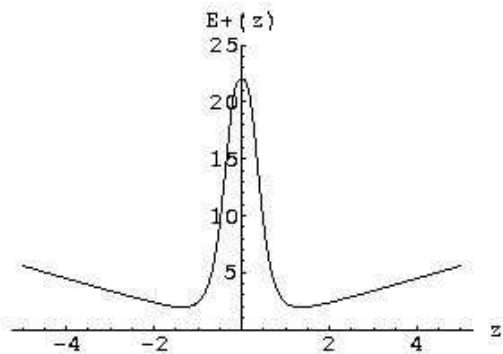}
\end{center}
\caption{\textbf{Potential energy barrier $E_+ ^R(z=z_1-z_2)$ after the
regularization.}}
\label{}
\end{figure}

Since the Hamiltonian (\ref{HBO}) commutes with the parity operator, the
eigenstates of it are also eigenstates of the parity, namely they are either
symmetric or antisymmetric. Let us denote these states by $|\phi
_{S}\rangle $ and $|\phi _{A}\rangle $ respectively. Then in general we
know $E_{S}\leq E_{A}$, where $E_{S}$ ($E_{A}$) is the eigenvalue of $|\phi
_{S}\rangle $ ($|\phi _{A}\rangle $). Now we consider a state in which the
particle is located on either the right side of the barrier or the left side
and express these states by $|\phi _{R}\rangle $ and $|\phi _{L}\rangle $
respectively. Notice that they are not eigenstates of the Hamiltonian (\ref%
{HBO}) in general. Obviously we have the following relation: 
\begin{eqnarray}
|\phi _{S}\rangle  &=&\frac{1}{\surd {2}}(|\phi _{R}\rangle +|\phi
_{L}\rangle ) \\
|\phi _{A}\rangle  &=&\frac{1}{\surd {2}}(|\phi _{R}\rangle -|\phi
_{L}\rangle )
\end{eqnarray}%
up to a phase factor. Therefore if we choose the initial state at $t=0$ as 
$|\phi _{R}\rangle $, then after simple calculations we get the state at
time $t$ as follows. 
\begin{equation} \label{psiR}
|\phi _{R}(t)\rangle \propto \cos (\frac{\Delta t}{\hbar })|\phi _{R}\rangle
+i\sin (\frac{\Delta t}{\hbar })|\phi _{L}\rangle ,
\end{equation}%
where $\Delta =(E_{A}-E_{S})/2$ and we omit an overall time dependent phase
factor which is irrelevant right now. Therefore we observe that the state is
oscillating between a configuration in which the particle is on the right
side and a configuration in which the particle is on the left side. In terms
of the original variables, the positions of two particles are being
exchanged in time.

To know the characteristic frequency $\Delta /\hbar $ we need to know the
eigenvalues and the states of the Hamiltonian using the relation $\Delta
=-\langle \phi _{R}|H_{BO}|\phi _{L}\rangle $. This requires knowledge
about the solution to the Schrodinger equation (\ref{schBO}).  Instead we
can use the WKB approximation to examine the possibility of this effect.

Using the WKB method we calculate the probability $w$ for a particle at rest
around one minima $r_{m}$ of the potential to tunnel across the potential
barrier to the other minimum: 
\begin{equation}
w(E=0)\simeq \exp \left[ -8\sqrt{\frac{m\mu ^{2}}{\hbar ^{2}}}(\frac{3}{%
\sqrt{r_{c}}}-\frac{2}{\sqrt{r_{m}}})\right] .  \label{w}
\end{equation}%
A cutoff distance of $10^{-15}m$ was chosen for $r_{c}$, approximately the
Compton wavelength of the neutron.   This value for $r_{c}$ and the value of 
$r_{m}$ from Appendix B yields an estimate for $w\sim e^{-0.94}\sim 0.39$.
It suggests a possibility of the tunneling effect, namely, the exchanging of
the two particles through the barrier. We summarize these calculations in
Appendix B.

Two remarks are in order: when the potential barrier becomes infinite the
ground state will be degenerate ($E_{S}=E_{A}\Leftrightarrow \Delta =0$).
This means there are no oscillations at all. And hence there exist the
states $|\phi _{R}\rangle $ and $|\phi _{L}\rangle $ separately. Although
these states are not eigenstates of the parity operator, because of the
degeneracy they are allowed states.  This is an example of the spontaneous
symmetry breaking. To examine the possibility of the exchanging effect
more precisely we also need to include the hyperfine interaction term in the
original potential (\ref{U}), which becomes important at short distances.


\section{Case 2: Magnetic field present on the right side of the origin only}

We assume the magnetic field $B(z)$ is in the $z$-direction and that it is
non-zero only in some region on the right side of the origin ($z_{1},z_{2}>0$%
). Outside of this region, for example on the left, the magnetic field
vanishes so, for example, $B(z_{1})\psi _{L}(z_{1})=0$, $B(z_{2})\psi
_{L}(z_{2})=0,$ etc. This magnetic field breaks the translational symmetry
of the field present in Case 1. Using these properties we can show that 
\begin{equation}
\Sigma _{+}\left\vert S\right\rangle =\Sigma _{+}\left\vert T\right\rangle =0
\end{equation}%
\begin{eqnarray}
\Sigma _{-}\left\vert S\right\rangle  &=&-B_{T}\left\vert T\right\rangle  \\
\Sigma _{-}\left\vert T\right\rangle  &=&-B_{T}\left\vert S\right\rangle 
\end{eqnarray}%
where we define 
\begin{equation}
B_{T}(z_{1},z_{2})=B(z_{1})+B(z_{2}).
\end{equation}%
Note that for our model, either particle 1 or 2 will be on the left side of
the origin so either $B(z_{1})$ or $B(z_{2})$ will vanish. Thus for the
special case of a constant field $B_{0}$ on the right side, we have $%
B_{T}=B_{0}.$ With our assumptions we can express $B_{T}(z_{1},z_{2})$ in
the following way: 
\begin{equation}
B_{T}(z_{1},z_{2})=B(z_{1})\theta (z_{1}-z_{2})+B(z_{2})\theta (z_{2}-z_{1})
\end{equation}

In the impulsive approximation, these results yield a representation of $%
H_{I}$ using the basis states $\{\left\vert T\right\rangle ,\left\vert
S\right\rangle \}$ 
\begin{equation}
H_{I}=\mu B_{T}\left[ 
\begin{array}{cc}
0 & 1 \\ 
1 & 0%
\end{array}%
\right] .
\end{equation}

We notice that the $B_{T}$ field causes transitions between the singlet and
triplet states like the inhomogeneous field in all space. This suggests
that even the simplest case (a constant field $B_{0}$ on the right side)
will do this. This transition might be of interest, in particular, in
applications of quantum computation. Similar discussion were done by
several authors (see Ref. \cite{unanyan} and references therein). However,
the transition between the singlet and the triplet was not discussed in there.

The remaining components of the triplet also transform among themselves,
which follows from: 
\begin{eqnarray}
\Sigma _{+}\left\vert T_{\pm 1}\right\rangle &=& \pm B_{T}\left\vert T_{\pm
1}\right\rangle \\
\Sigma _{-}\left\vert T_{\pm 1}\right\rangle &=&0.
\end{eqnarray}%
The interaction Hamiltonian in the basis \{$\left\vert T_{-1}\right\rangle
,\left\vert T_{1}\right\rangle $ \} is: 
\begin{equation}
H_{I}=\mu B_{T}\left[ 
\begin{array}{cc}
1 & 0 \\ 
0 & -1%
\end{array}%
\right]
\end{equation}%
In terms of the basis $\{ \left\vert T_{-1}\right\rangle ,\left\vert
T_{1}\right\rangle ,\left\vert T\right\rangle ,\left\vert S\right\rangle \}$
, the total reduced Hamiltonian $H^{R}$ can therefore be written as 
\begin{equation}
H^{R}=f(r)\left[ 
\begin{array}{cccc}
-2 & 0 & 0 & 0 \\ 
0 & -2 & 0 & 0 \\ 
0 & 0 & 4 & 0 \\ 
0 & 0 & 0 & 0%
\end{array}%
\right] +\mu B_{T} \left[ 
\begin{array}{cccc}
1 & 0 & 0 & 0 \\ 
0 & -1 & 0 & 0 \\ 
0 & 0 & 0 & 1 \\ 
0 & 0 & 1 & 0%
\end{array}%
\right] .
\end{equation}

After applying the magnetic field perturbation, the $\left\vert
T_{-1}\right\rangle $ and $\left\vert T_{1}\right\rangle $ components still
do not mix with any other components. The total reduced Hamiltonian is
diagonal in the subspace spanned by $\left\vert T_{-1}\right\rangle $ and $%
\left\vert T_{1}\right\rangle ,$ and the corresponding energy eigenvalues
are $E_{-1}^{R}=-2f+\mu B_{T}$ and $E_{1}^{R}=-2f-\mu B_{T}$.

We now consider in detail the subspace spanned by $\left\vert T\right\rangle
,\left\vert S\right\rangle $ which is not diagonal. In the same manner as
in the previous section we express the total reduced Hamiltonian $H_{2}^{R}$
in terms of the Pauli matrices as: 
\begin{equation}
H_{2}^{R}=2f(r)\bm{I}+\bm{\alpha }\cdot \bm{\sigma }
\end{equation}%
where the vector $\bm{\alpha }$ $=(\mu B_{T},0,2f)$. Define the angle $%
\theta $ by 
\begin{equation}
\tan \theta =\frac{\alpha _{x}}{\alpha _{z}}=\frac{\mu B_{T}(z_{1},z_{2})}{2f%
}=\frac{B_{T}r^{3}}{2\mu }
\end{equation}%
and the quadrant is specified by%
\begin{equation}
\sin \theta =\frac{\mu B_{T}}{\sqrt{\mu ^{2}B_{T}^{2}+4f^{2}}}.
\end{equation}%
Solving for the eigenvectors and eigenvalues of $H_{2}^{R}$, we obtain 
\begin{eqnarray}
\left\vert -a\right\rangle  &=&-\sin \frac{\theta }{2}\left\vert
T\right\rangle +\cos \frac{\theta }{2}\left\vert S\right\rangle , \\
E_{-}^{R} &=&2f-|\bm{ \alpha }|=2f(1-\sec \theta )
\end{eqnarray}%
and 
\begin{eqnarray}
\left\vert +a\right\rangle  &=&\cos \frac{\theta }{2}\left\vert
T\right\rangle +\sin \frac{\theta }{2}\left\vert S\right\rangle , \\
E_{+}^{R} &=&2f+|\bm{ \alpha }|=2f(1+\sec \theta ).
\end{eqnarray}

To avoid the crossing of energy levels, we require $E_{-}^{R}>E_{-1}^{R}$
which implies that%
\begin{equation}
2\mu B_{T}<3f(r).  \label{levelcond}
\end{equation}%
And also from the adiabatic theorem we get 
\begin{equation}
T\gg \frac{\hbar (\mu B_{T})^{2}}{(4f)^{2}\sqrt{(\mu B_{T})^{2}+(2f)^{2}}}.
\end{equation}%
Using the above condition (\ref{levelcond}) we estimate the lower limit of
the time $T_{min}$ as 
\begin{equation}
T_{min}\sim \frac{9}{80}\frac{\hbar }{f(r)}\simeq 0.11\hbar f^{-1}.
\label{adiabaticcond}
\end{equation}

The restriction of Eq. (\ref{levelcond}) severely limits possible modulation
frequencies of the field being applied adiabatically. We briefly mention a
procedure, similar to that used in Case 1, which results in a much larger
bandwidth. We first apply the large constant $B_{0}$ field impulsively
over all space \cite{impulsive}. Then we apply the inhomogeneous field
adiabatically over the right side only. With this procedure, the
restriction on the energy levels to avoid crossings is similar to Eq. (\ref%
{equality1}), but the levels $|T_{-1}\rangle $ and $|T_{+1}\rangle $ are
also shifted by the inhomogeneous field $\pm \mu B_{T}|_{B_{0}=0}$.  The
quantity $\mu B_{T}|_{B_{0}=0}$ is always positive since the inhomogeneous
field is on the right side, so the requirement for no level crossing is less
stringent, and is met simply if $\mu B_{0}>4f$. The corresponding
requirement for the validity of the adiabatic approximation is the same as
Eq. (\ref{adia1}), with $r$ replaced by the positive value of the pair $%
\{z_{1},z_{2}\}$.

\subsection{Homogeneous field on the right side only}

We can consider several different magnetic field strengths $%
B_{T}(z_{1},z_{2}).$ \ For the simplest case, in which $B_{T}$ is a constant
on the right side, we find that $E_{-}^{R}(z_{1},z_{2})$ is a negative
potential (Figures 6--7) that does not confine the particles so no stable
states are expected, while $E_{+}^{R}(z_{1},z_{2})$ is a positive potential
that also does not confine particles (Figures 8--9).

\begin{figure}[htbp]
\begin{center}
\includegraphics[width=3in]{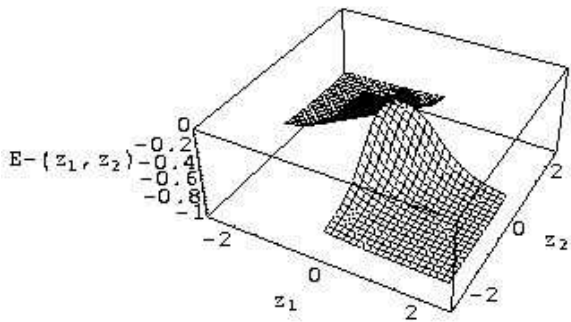}
\end{center}
\caption{\textbf{$E_- ^R(z_1,z_2)$ for a constant magnetic field on the
right side, plotted for particles on opposite sides of the origin.}}
\label{}
\end{figure}

\begin{figure}[htbp]
\begin{center}
\includegraphics[width=3in]{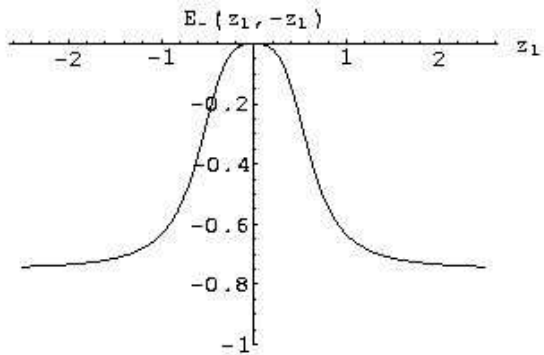}
\end{center}
\caption{\textbf{$E_- (z1,-z1)$ for a constant magnetic field This curve is
obtained by plotting the function $E_-(z1,z2)$ for the contour $z1+z2=0$ .}}
\label{}
\end{figure}

\begin{figure}[htbp]
\begin{center}
\includegraphics[width=3in]{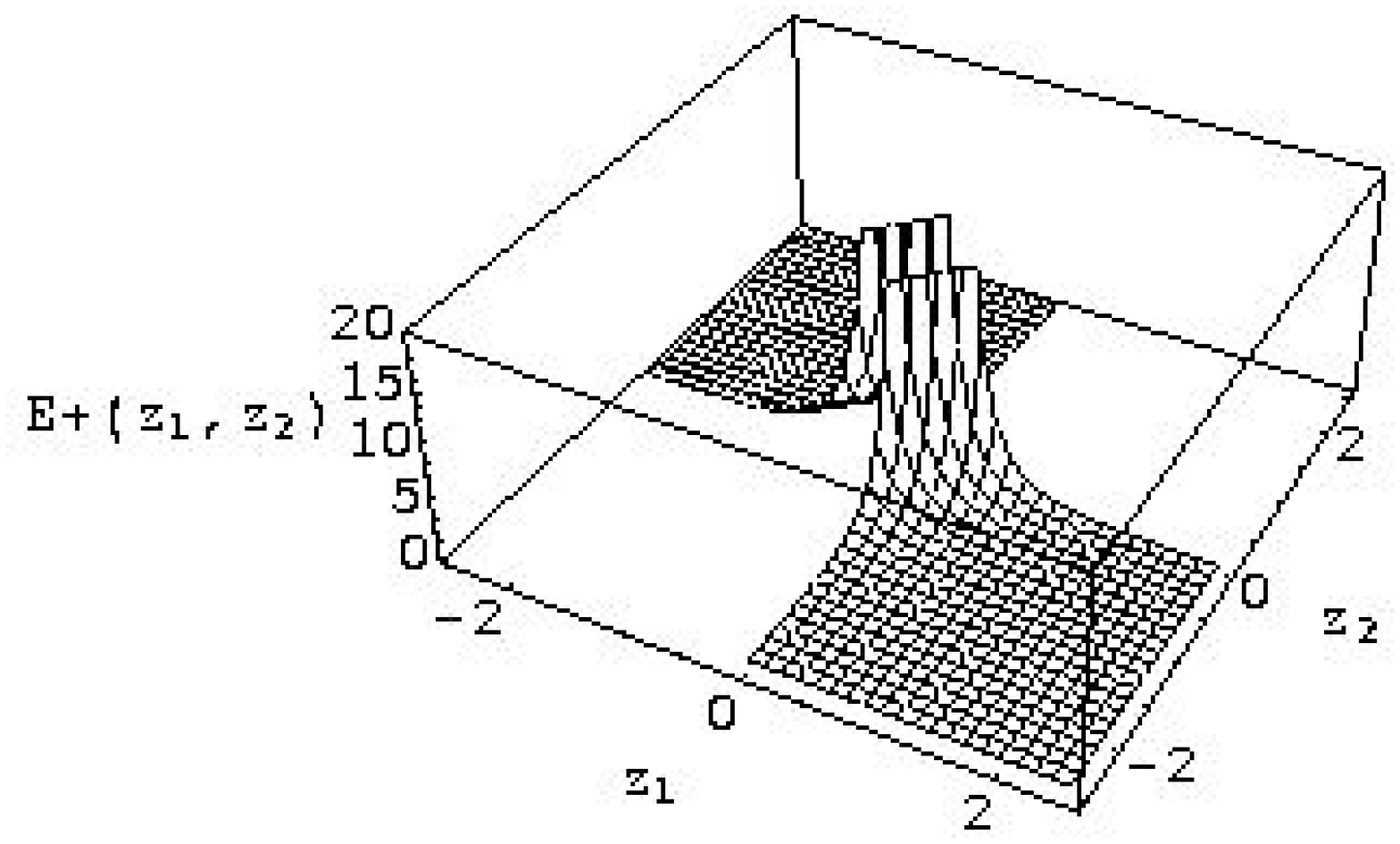}
\end{center}
\caption{\textbf{$E_+(z1,z2)$ for a constant magnetic field on the right,
plotted only for particles on opposite sides of the origin.}}
\label{}
\end{figure}

\begin{figure}[htbp]
\begin{center}
\includegraphics[width=3in]{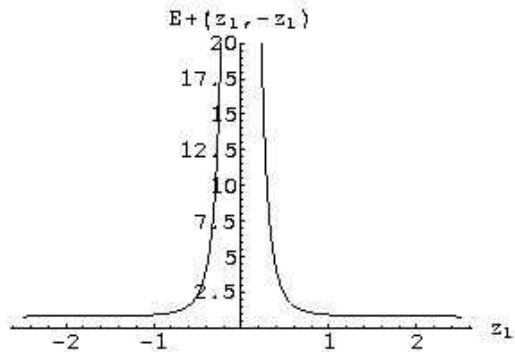}
\end{center}
\caption{\textbf{$E_+(z1,-z1)$ plotted for a constant magnetic field on the
right side of the origin.}}
\label{}
\end{figure}

\subsection{Inhomogeneous field on the right side only}

Next we shall consider an inhomogeneous field ($B_{T}=B_{0}+bZ+br/2$) on the
right side. The graphs for $E_{-}^{R}(z_{1},z_{2})$ (FIG. 10--11) are
plotted with the same parameters as FIG. 1--4 for the comparison. Similar
results are obtained in this case.

The positive energy $E_{+}^{R}(z_{1},z_{2})$ has a double hump structure
(FIG. 12--13) with two potential wells for quasi-bound states separated by
an energy peak. We note that unlike in case 1, there is no symmetry with
respect to reflection across the line $z_{1}+z_{2}=0$ so the forces acting
on the particles are not equal. The force due to the magnetic field acts
only on the particle on the right side. Here we define the effective
forces through partial differentiations of the effective potentials $E_{\pm
}^{R}$ with respect to $z_{1}$ and $z_{2}$. 
\begin{figure}[tbph]
\begin{center}
\includegraphics[width=3in]{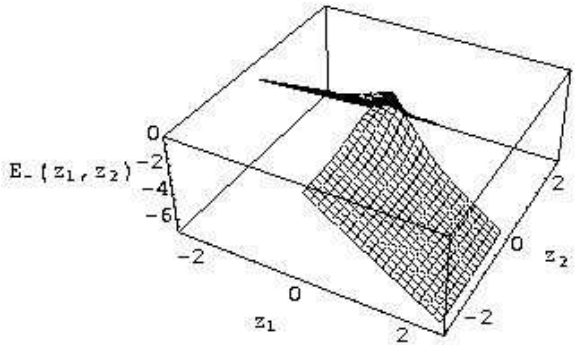}
\end{center}
\caption{\textbf{$E_{-}(z1,z2)$ for an inhomogeneous field on the right
side, plotted for particles on opposite sides of the origin. All parameters
for the FIG. 10--13 are chosen same as for FIG. 1--4.}}
\end{figure}

\begin{figure}[htbp]
\begin{center}
\includegraphics[width=3in]{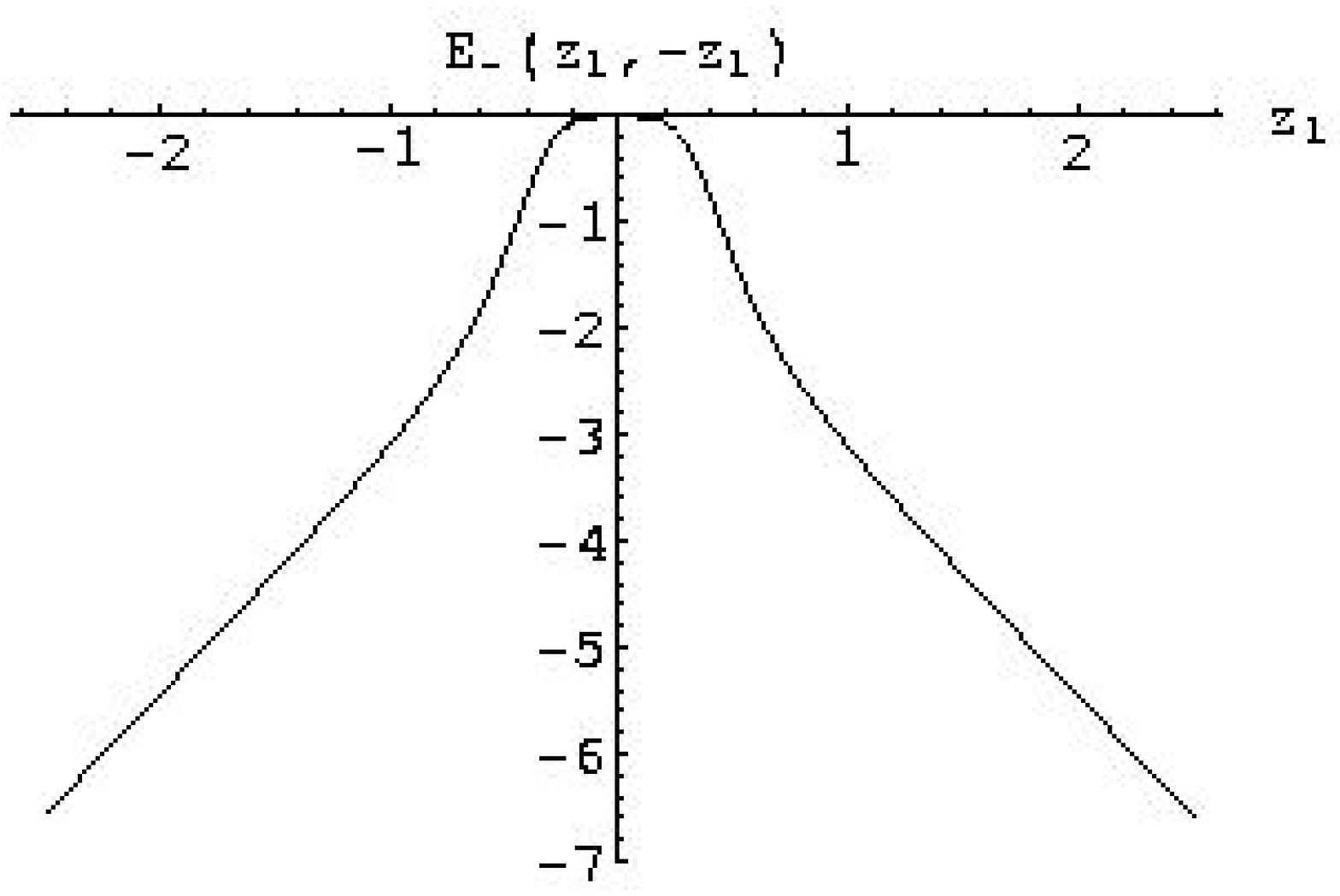}
\end{center}
\caption{\textbf{$E_-(z1,-z1)$ for an inhomogeneous field on the right side
only.}}
\label{}
\end{figure}

\begin{figure}[htbp]
\begin{center}
\includegraphics[width=3in]{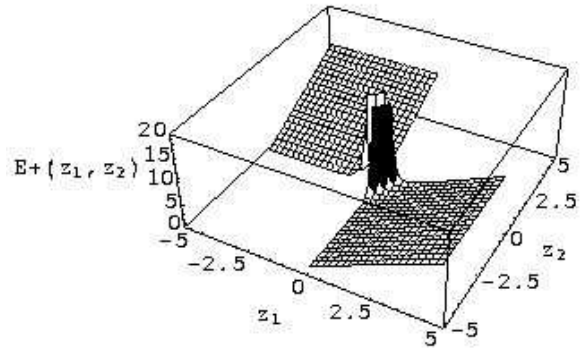}
\end{center}
\caption{\textbf{$E_+(z1,z2)$ for an inhomogeneous field on the right side,
plotted for particles on opposite sides of the origin. }}
\label{}
\end{figure}

\begin{figure}[htbp]
\begin{center}
\includegraphics[width=3in]{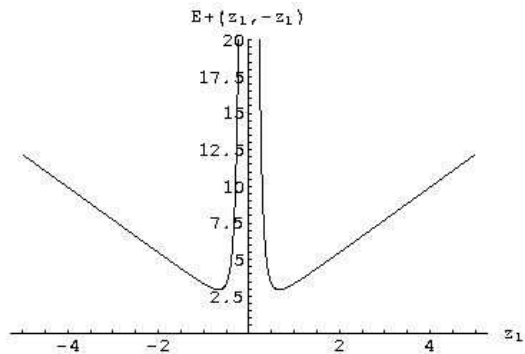}
\end{center}
\caption{\textbf{$E_+(z1,-z1)$ for an inhomogeneous field on the right side. 
}}
\label{}
\end{figure}

\subsection{Summary}

In summary, with the adiabatic evolution of spins when we apply the magnetic
field on the right side, we find that the initial singlet ground state
singlet $\left\vert S\right\rangle $ has evolved into the eigenstate given
by a linear combination of singlet and triplet states: 
\begin{equation}
\left\vert -a\right\rangle =-\sin \frac{\theta }{2}\left\vert T\right\rangle
+\cos \frac{\theta }{2}\left\vert S\right\rangle .
\end{equation}%
In this adiabatic evolution, the reduced energy goes from $0$ for $%
\left\vert S\right\rangle $ to $2f(1-\sec \theta )$ for $\left\vert
-a\right\rangle $. Since $\left\vert -a\right\rangle $ is a spin
eigenstate of the total Hamiltonian, there is an overall time dependent
phase factor, $e^{-iE_{-}^{R}t/\hbar }$ for this state, which we have
omitted.

The eigenstates $\left\vert \pm a\right\rangle $ have the property that, for
any value of $\theta $, the spins of the particles in the $z$-direction are
always correlated: 
\begin{equation}
(S_{z_{1}}+S_{z_{2}})\left\vert \pm a\right\rangle =0.
\end{equation}%
This result, which follows since $[H_{T},S_{z_{1}}+S_{z_{2}}]=0$, indicates
that the spins of the two particles remain correlated if an inhomogeneous
magnetic field is applied only to the particle on the right side of the
origin.


\section{Computation of the density matrix for an adiabatic change in the
magnetic field on the right side only}

In this section we shall study the case 2 further in detail. One convenient
way to consider the effect of the magnetic field (which is present only on
the right side) on the state of the system on the left side (where the
magnetic field vanishes) is to predict the results of a measurement of an
observable $A$ which is defined only on the left side. The predictions can
be done by means of the reduced density matrix $\rho (L,-a)$ on the left
side $L$ for the state $\left\vert -a\right\rangle $\cite{density}:

\begin{equation}
\rho (L,-a)=Tr_{right\text{ }side}\left\vert -a\right\rangle \left\langle
-a\right\vert .
\end{equation}%
In order to compute this reduced density matrix for $\left\vert
-a\right\rangle $, we first compute the reduced density matrices for the
basis states $\left\vert S\right\rangle $ and $\left\vert T\right\rangle $.
\ For an observable $A$ which is defined only on the left side we have $%
A\psi _{R}(z_{2})=0$, etc. The result of a measurement of $A$ for the state $%
\left\vert T\right\rangle $ is defined as 
\begin{equation}
Tr\rho (L,T)A=\left\langle T\right\vert A\left\vert T\right\rangle 
\end{equation}%
and can be computed using the expression for $|T\rangle $: 
\begin{eqnarray}
\langle T|A|T\rangle  &=&\frac{1}{4}[{}_{1}\langle +L|A|+L\rangle
_{1}+{}_{1}\langle -L|A|-L\rangle _{1} \\
&&\ +{}_{2}\langle -L|A|-L\rangle _{2}+{}_{2}\langle +L|A|+L\rangle _{2}]
\end{eqnarray}%
where we have defined 
\begin{eqnarray}
\left\vert \pm L\right\rangle _{1} &\equiv &\left\vert \pm \right\rangle
_{1}\psi _{L}(z_{1}) \\
\left\vert \pm L\right\rangle _{2} &\equiv &\left\vert \pm \right\rangle
_{2}\psi _{L}(z_{2}).
\end{eqnarray}%
The corresponding reduced density matrix on the left side is 
\begin{eqnarray}
\rho (L,T) &=&\frac{1}{4}[|+L\rangle _{1}{}_{1}\langle +L|+|-L\rangle
_{1}{}_{1}\langle -L| \\
&&\ +|+L\rangle _{22}\langle +L|+\text{\ }|-L\rangle _{22}\langle -L|].
\end{eqnarray}

By a similar calculation we find that the reduced density matrix for $\left|
S\right\rangle $ has the same value as for $\left| T\right\rangle $:%
\begin{eqnarray}
\rho (L,S) &=&Tr_{rightside}\left| S\right\rangle \left\langle S\right| \\
&=&\rho (L,T).
\end{eqnarray}

We also compute 
\begin{eqnarray}
\rho (L,TS) &=&Tr_{rightside} | T \rangle \langle S | \\
&=&\frac{1}{4} [ | +L \rangle _1 {}_1 \langle +L | - | -L \rangle _1 {}_1
\langle -L | \\
& & \ + | +L \rangle _2 {}_2 \langle +L | - | -L \rangle _2 {}_2 \langle -L
| ]
\end{eqnarray}

and find that 
\begin{equation}
\rho (L,TS)=\rho (L,ST).
\end{equation}

Using the preceding results, we can write an expression for $\rho (L,-a)$
using the definition of the state $\left| -a\right\rangle$:%
\begin{eqnarray} \nonumber
\rho (L,-a) &=&\sin ^{2}\frac{\theta }{2}\rho (L,T)+\cos ^{2}\frac{\theta }{2%
} \rho (L,S) \\
&\quad & -2\sin \frac{\theta }{2}\cos \frac{\theta }{2}\rho (L,TS) \\
&=&\rho (L,S)- \rho (L,TS) \sin \theta .
\end{eqnarray}

This last result can be written in matrix form in the basis \{$\left|
+L\right\rangle _{1},\left| -L\right\rangle _{1},\left| +L\right\rangle
_{2},\left| -L\right\rangle _{2}\}:$%
\begin{equation}
\rho (L,-a)= \frac 14 \left[ 
\begin{array}{cccc}
1-\sin \theta & 0 & 0 & 0 \\ 
0 & 1+\sin \theta & 0 & 0 \\ 
0 & 0 & 1-\sin \theta & 0 \\ 
0 & 0 & 0 & 1+\sin \theta%
\end{array}%
\right].
\end{equation}

One can contemplate making standard quantum mechanical measurements in which
a superposition collapses to a eigenstate\cite{peresbook}. Alternatively
we consider the use of a protective Stern-Gerlach measurement, in which
there is an adiabatic interaction between the pointer and the system\cite%
{av,aavwave,aavprotec}. We want to examine the possibility of the protective
measurement in our model under the assumption we will work in a decoherence
free subspace. In previous discussions of protective measurements, only
the case of a small perturbation was considered in order to minimize the
change in the wavefunction of the state to be measured. In our system, such
restrictions are not necessary. As long as the interaction is adiabatic,
the state will evolve continuously as an eigenstate of the instantaneous
Hamiltonian, without any transition provided there is no degeneracy in the
energy. If the magnetic field is applied adiabatically, the field can
become large, resulting in a large change in the wavefunction. Because
the changes are adiabatic, they are reversible as the magnetic field is
reduced. Since a protective measurement of the spin for our two particle
system does not change the state of the system, the protective measurement
does not end the entanglement.

In the spin-spin model, the states $\left\vert -a\right\rangle $ and $%
\left\vert +a\right\rangle $ into which $\left\vert S\right\rangle $ and $%
\left\vert T\right\rangle $ evolve when a magnetic field is applied on the
right side are non-degenerate so a protective measurement of the reduced
density matrix should be possible. Indeed all four basis states$\left\vert
-a\right\rangle $, $\left\vert +a\right\rangle $, $\left\vert
T_{1}\right\rangle $, and $\left\vert T_{-1}\right\rangle $ are
non-degenerate provided $f\neq 0$ and $B\neq 0$ so a protective measurement
should be possible of the density matrix\cite{anandanden}. As a
consequence the result of a protective measurement of an observable $A$ that
acts only on the left side can be written as the trace over the reduced
density matrix:%
\begin{equation}
<A>_{(L,-a)}=Tr(\rho (L,-a)A).
\end{equation}

Note that here $<A>$ represents the result of a single protective
measurement on a single system. It does \textbf{not} have the usual meaning
of the expectation value of $A$, which is based on the measurement of the
observable $A$ for an ensemble of identically prepared systems. Since the
elements of the density matrix depend on $\theta $, it is clear that changes
in $\theta $, resulting from either changes in $B$ or ($z_{1}-z_{2})$, will
affect the measured values of observables $A$. In other words, in this
model with a finite separation, by changing the value of the magnetic field
on the right side, it is possible to detect the effect on the left side
nonlocally. Note that nowhere do we use the specific form of the
potential. Indeed, we could use the formalism presented with any function
of the separation.

To illustrate the application of a protective measurement, we could measure
the total spin in the z- direction for both particles using local
measurements on the left side only. We assume we use a Stern-Gerlach
analyzer with an inhomogeneous field in the $z$-direction and we observe the
deflection of the wave packet corresponding to a particle. We assume we
have calibrated our apparatus so that we can determine the spin of a
particle on the left side by observation of the deflection in the $z$-
direction at a certain point in the experiment. Since the particles are
indistinguishable, a single protective measurement of the spin of a particle
on the left will yield a deflection corresponding to the result 
\begin{equation}
\left\langle S_{z_{1}}+S_{z_{2}}\right\rangle _{left\ side}=tr(\rho \lbrack
S_{z_{1}}+S_{z_{2}}])=-\frac{1}{2}\sin \theta 
\end{equation}%
where the value of $\theta $ is determined by the magnetic field on the
right side and the distance between the particles. The corresponding total
spin in the $z$-direction measured \textbf{on the right side} is $\frac{1}{2}$%
sin$\theta $. Since this protective measurement has not collapsed the wave
function, we could do additional measurements \textbf{on the same state} to
determine the rest of the density matrix $\rho _{(L,-a)}$ in the $\left\vert
T\right\rangle ,\left\vert S\right\rangle $ subspace. If we used
Stern-Gerlach analyzers with fields in the $x$ and $y$ directions for two
additional experiments, respectively, we would find: 
\begin{equation}
\left\langle S_{x_{1}}+S_{x_{2}}\right\rangle _{left\ side}\ =\left\langle
S_{y_{1}}+S_{y_{2}}\right\rangle _{left\ side}=0.
\end{equation}%
These components of the spin vanish because we chose the magnetic field on
the right side to be in the $z$-direction.

It is interesting to contrast the protective measurement of $%
S_{z_{1}}+S_{z_{2}}$ with the ordinary measurements. In an standard quantum
mechanical measurement the wavefunction collapses and an eigenvalue of the
operator is measured. To determine the expectation value of the operator,
one needs to make a statistically significant number of measurements on an
ensemble of identically prepared systems. In our model, since the
particles are indistinguishable, for each standard measurement on a
different identically prepared system, a deflection in the Stern-Gerlach
apparatus will be measured that corresponds to spin of $+\frac{1}{2}$ or $-%
\frac{1}{2}$. If enough measurements are done, it will be determined that
the probability of measuring $+\frac{1}{2}$ is $\frac{1}{2}(1-\sin \theta )$
and the probability of measuring $-\frac{1}{2}$ is $\frac{1}{2}(1+\sin
\theta )$. Using these probabilities, the standard measurements will
determine that the expectation value of the operator $S_{z_{1}}+S_{z_{2}}$
on the left side is $-\frac{1}{2}\sin \theta ,$ in agreement with the 
\textit{single} protective measurement.

This comparison of the two methods illustrates some of the advantages of
protective measurements and some of the disadvantages of standard
measurements.


\section{CONCLUSION AND DISCUSSIONS}

A system was considered in which two neutral spin 1/2 particles interact
through a diple-dipole potential. The potential leads to singlet and triplet
entangled states which are modified when we apply a magnetic field over all
space (Case 1), or over just the region to the right of the origin (Case 2).

For the case 1 we showed that the singlet state is essentially unstable and
it tends to separate. However, the spin zero component of the triplet will
have a double hump shape energy which implies the particles will tend to
stay near the minima with high probabilities and form a bound state. Also
we showed that it is possible to have a tunneling effect in this case. 
Such an effect might have applications in quantum computation.

For the case 2 we observed that although the external field is applied only
in one part of the system, the other part will be affected by it due to the
entanglement of the system. This manifests the nonlocality of the
entanglement in quantum mechanics. It appears protective measurements can
be used to determine the density matrix without ending the entanglement. 
As discussed above for the case of finite separation, one limitation in the
signaling between the two regions of space in this model lies in the
requirement that we must have an adiabatic perturbation or transitions will
occur between states. When the magnetic field is turned on, it must be
done slowly enough so that no transitions are induced between the initial
state and other states. Although the specific restrictions depend on the
manner in which the perturbations are applied, they require that the
potential due to the dipole-dipole interaction does not vanish.

\begin{acknowledgments}
We are deeply saddened to report that Jeeva Anandan died during the
preparation of this paper. This work was partially supported by the NFS,
grant No. PHY-0140377. GJM would like to thank the NASA Institute for
Advanced Concepts for supporting this work.
\end{acknowledgments}

\appendix 

\section{Model with confining potentials}

In this appendix we will examine our model with additional confining
potentials (for instance, created by the optical laser). For simplicity we
approximate this potential by a harmonic potential around certain point $%
\bm{x}_0$ with characteristic frequency $\Omega$, namely, $V(\bm{x})=\frac
12 m \Omega ^2 (\bm{x}-\bm{x}_0)^2$. We choose the center of the potential
in such a way that two particles will be separated by a distance $z_0$.
Under these conditions the ground state energy of the particles is $\frac 32
\Omega$ (we use $\hbar =1$ in appendices.) and the corresponding states are 
\begin{eqnarray}  \label{R}
\psi_R (\bm{x}_i) &=& (\frac{\xi}{\pi})^{\frac 34} e^{-\frac12 \xi (x_i
^2+y_i^2+(z_i-\frac{z_0}{2})^2)} \\
\psi_L (\bm{x}_i) &=& (\frac{\xi}{\pi})^{\frac 34} e^{-\frac12 \xi (x_i
^2+y_i^2+(z_i+\frac{z_0}{2})^2)} ,  \label{L}
\end{eqnarray}
for $i=1,2$ and the parameter $\xi = \sqrt{m \Omega}$ is introduced. To have
a stable ground state in our model we require the energy levels are well
separated, or the parameter $\xi$ is large. (Similarly we can assign these
wave packets for the free particles case too.) After we obtain the spatial
wave functions for the singlet and the triplets \cite{comment2}, we can
estimate the effect of the hyperfine interaction term in the dipole-dipole
potential (\ref{U}). We evaluate the expectation values of the delta
function as follows. 
\begin{eqnarray}  \label{dex}
\langle \delta (\bm{x}) \rangle _S&=&4(\frac{\xi}{2 \pi})^{\frac 32}
(1+e^{\frac 12 \xi z_0^2})^{-1}, \\
\langle \delta (\bm{x}) \rangle _T&=&0 .
\end{eqnarray}
Therefore, under our assumptions the contribution from the hyperfine
interaction is negligible in our model because of the exponential factor in (%
\ref{dex}). However, in contrast, this term plays an important role in a
situation where the distance becomes very small and we need to add this
contribution too. This regime is discussed in the literature, for
instance, see Ref. \cite{you} and references therein.

Next we estimate the change in the positions of the particles and the
kinetic energy of them. From simple calculations it is easy to see that the
expectation values of the relative position and the center of mass are zero.
Similarly we obtain for the expectation value for the center of mass
kinetic energy 
\begin{equation}
\left\langle \frac{P^2}{2m} \right\rangle _S=\left\langle \frac{P^2}{2m}
\right\rangle _T= \frac{\xi}{2m}
\end{equation}
and for the relative kinetic energy we have 
\begin{eqnarray}
\left\langle \frac{p^2}{m} \right\rangle _S &=& \frac{\xi}{4m} (1-\frac{\xi
z_0^2}{e^{\frac 12 \xi z_0^2}+1}) \\
\left\langle \frac{p^2}{m} \right\rangle _T &=& \frac{\xi}{4m} (1+\frac{\xi
z_0^2}{e^{\frac 12 \xi z_0^2}-1}) .
\end{eqnarray}
Again, the second term inside the parenthesis is negligible by our
assumptions. And hence we see that the corrections due to the motion of the
particles are small.

\section{Calculation of the tunneling probability (\ref{w})}

To estimate the tunneling probability we simplify our model as follows.
First we calculate the minima $r_m$ of the potential (\ref{VB}), this is
easily found 
\begin{equation}
gr_m=2 \sqrt{15} f(r_m) \Leftrightarrow r_m= (\frac{240 \mu ^2}{b^2})^{\frac
18} .
\end{equation}
Then we evaluate the the tunneling probability between two minima through
the dipole-dipole interaction barrier assuming that the potential is mainly
dominated by this interaction in small distance. We use the WKB method \cite%
{landau} to get the probability $w(E)$ for the particle with the kinetic
energy $E=k^2/m$ (the mass is $m/2$), that is, 
\begin{equation}  \label{w1}
w(E) \simeq \exp [-2 \left| \int _{-r_m}^{r_m} \sqrt{m (V(z)-E)} \ dz
\right| \ ] .
\end{equation}
In our case the simplified potential is $V(z)=4\mu ^2 /|z|^3$. As we
mentioned earlier we introduce the cutoff at $r_c$ to get rid of the
divergent integral. Under these assumptions and defining the momentum $k_m$
and $k_c$ corresponding to the minima and the cutoff respectively, i.e. $%
k_m^2/m=V(r_m)$ and $k_c^2/m=V(r_c)$, the integral inside the exponent of (%
\ref{w1}) is evaluated as below. 
\begin{equation}
-2 \sqrt{k_m^2-k^2}\ r_m -k r_m (\frac{k_m}{k})^{\frac 13} B_{\frac 56
,\frac 12} (\frac{k}{k_m}) +3 \sqrt{k_c^2-k^2}\ r_c ,
\end{equation}
where $B_{a,b} (x)$ is the incomplete beta function defined by $B_{a,b}
(x)=\int _0 ^x t^{a-1}(1-t)^{b-1} dt $ and we neglect the positive orders of 
$r_c$ (in $r_c/r_m \ll 1$ limit). Therefore for the rest particle ($k=0$) at
the minima, a rough estimate for the tunneling probability is 
\begin{equation}  \label{E0}
w(E=0) \simeq \exp [-4 (3 k_c r_c -2 k_m r_m) ]
\end{equation}
which is given in the text after substituting $k_m$ and $k_c$. Another case
is for the particle having the minimum potential energy ($k=k_m$): 
\begin{equation}
w(E=V(r_m)) \simeq \exp [-4 (3 k_c r_c -B(\frac 56, \frac 12) k_m r_m) ]
\end{equation}
where $B(\frac 56, \frac 12)\simeq 2.24$ is the usual beta function. And
hence we see that this is the same order as (\ref{E0}).



\begin{thebibliography}{99}
\bibitem{comment1} Strictly speaking the magnetic field must satisfy the
Maxwell equations, i.e., $\mathrm{div}\bm{B}=0$ and $\mathrm{rot}\bm{B}=0$
(without external currents). In our case the simplest realistic
inhomogeneous field might be chosen as $\bm{B}=(0,-by,B_{0}+bz)$ which
satisfies the Maxwell equations. This choice of the field and the assumption 
$|by|\ll B_{0}$ yields the same result as obtained by our choice $\bm{B}%
=(0,0,B_{0}+bz)$ (\ref{B1}). See also the discussions in Ref. \cite{anandandivB}
\bibitem{commentBb} This application of the field eliminates the non-level
crossing condition (\ref{equality1}). On the other hand, if we apply both
parts of the magnetic field adiabatically, we will have more restrictive
condition which limits the capabilities of our model.
\bibitem{super} We can imagine two identical particles moving toward each
other which are represented by Gaussian wave packets which are
superpositions of the different momenta of the plane waves. The time
development of this system is similar to the one discussed in the text after eq. (\ref{psiR}).
\bibitem{condition} This sufficient condition is obtained by using the
inequality $|x|+|y|\geq \sqrt{x^{2}+y^{2}}$ and $Z\approx 0$.
\bibitem{impulsive} The validity of this impulsive approximation is
guaranteed by an equality $T_{imp}\ll \frac{2\hbar }{\mu B_{0}}$ \cite%
{messiah}.
\bibitem{density} Similarly we can evaluate the reduced density matrix for the state $|+a \rangle$ 
and it is found to be $\rho(L,+a) =1/4 \ {\rm diag}(1+\sin \theta ,1-\sin \theta ,1+\sin \theta ,1-\sin \theta )$.
\bibitem{comment2} The spatial wave functions for the singlet and the
triplets would not be normalized by the factor $1/ \sqrt{2}$ if we use these
wave packets (\ref{R},\ref{L}). This is because there is overlapping of two
wave packets. And we need to renormalize the spatial wave functions after
forming (\ref{SA}). Therefore these wave packets are approximated trial wave
functions.
\end{thebibliography}
\end{document}